\documentclass{PoS}

\title{Flux variability in the FSRQ 1510-089. A radio-gamma perspective}

\ShortTitle{Flux variability in the FSRQ 1510-089}

\author{\speaker{M. Orienti}\thanks{This research has made use of the
    data from the MOJAVE database that is maintained by the MOJAVE
    team (Lister et al. 2009, AJ, 137, 3718)}, G. Giovannini, D. Dallacasa\\
        Astronomy Department, Bologna University, Italy\\
        E-mail: \email{orienti@ira.inaf.it}}
\author{M. Giroletti, T. Venturi\\
        INAF-IRA Bologna, Italy}
\author{F. D'Ammando, S. Vercellone\\
        INAF-IASF Palermo, Italy}
\author{M. Tavani\\
        INAF-IASF Roma, Italy}

\abstract{The knowledge of the physical conditions occurring in the
  relativistic jet of radio-loud active galactic nuclei (AGNs) is
  important to understand the mechanisms at the basis of their
  multiband emission. From parsec-scale radio observations of blazar
  objects it has been suggested a connection between the ejection of
  new jet features and strong $\gamma$-ray flares. We present results
  from multi-epoch Very Long Baseline Interferometer (VLBI) and
  Space-VLBI observations of the Flat Spectrum Radio Quasar (FSRQ)
  PKS\,1510-089. The comparison of the radio structure observed at 
  different epochs shows the presence of jet features moving with 
  highly superluminal apparent velocity. 
  Radio flux density variability and changes in the
  source structure and in the polarization properties 
  are then compared with the information on the
  $\gamma$-ray emission in order to find a possible connection between
radio and $\gamma$-ray emission.}

\FullConference{10th European VLBI Network Symposium and EVN Users Meeting: VLBI and the new generation of radio arrays\\
		September 20-24, 2010\\
		Manchester Uk}

\begin{document}

\section{Introduction}

Only a small fraction of the AGN population display a powerful radio
emission. This phenomenon is associated with relativistic particles
produced in the central region of the 
AGN and channelled through the jets towards
the outermost regions. In addition to the synchrotron radiation
detected in the radio band, the relativistic particles may be also
responsible for the high-energy emission by means of scattering of
low-energy photons, suggesting in this way a connection between radio
and $\gamma$-ray emission. The fact that all the AGN detected by EGRET
were powerful radio-loud objects strongly support a relation between
radio and $\gamma$-ray emission. However, such a relation has not been
proved unambiguously. The class of AGN dominating the $\gamma$-ray sky
is the blazar population. These objects are characterized by the
presence of a compact radio core, superluminal jets, and extreme flux
density and polarization variability. Their properties are interpreted
as the result of severe beaming effects caused by the orientation of
the jet axis at very small angles with the line of sight. Among the
blazar population, the flat spectrum radio quasar PKS 1510-089
($z$=0.361) is the ideal target to locate the region responsible for
the emission. Enhanced
$\gamma$-ray activities have been detected several times 
by both AGILE and {\it Fermi}
satellites \cite{abdo10,dammando09,pucella08}. In the radio band the
source is characterized by highly superluminal jet knots with
apparent velocity of about 20$c$ \cite{homan02}. Thanks to the very
small angle of a few degrees formed by the jet with the line of sight,
PKS\,1510-089 represents the good candidate to investigate the
possible connection between radio and $\gamma$-ray emission. \\

\section{The source morphology}
To study the parsec-scale morphology, PKS\,1510-089 was target of
multi-epoch Space-VLBI observations at 5 GHz and VLBA observations at
8.4 GHz. The excellent spatial resolution ($\sim$
1.8$^{\prime\prime}$$\times$0.9$^{\prime\prime}$ at 5 GHz and
$\sim$2.3$^{\prime\prime}$$\times$1.0$^{\prime\prime}$ at 8.4 GHz) 
allowed us to accurately describe the source structure and to follow
the ejected knots during their motion. Additional multi-epoch 
VLBA 15-GHz data from the MOJAVE programme \cite{lister09}
were analysed in order to
complement the morphological information with data at a higher frequency
and to study on a longer time baseline the flux density and
polarization variability of the core component. More details on
the observations and the data analysis can be found in \cite{orienti10}.\\
Dual-epoch Space-VLBI observations at 5 GHz (Fig. 1a) 
show that the parsec scale
radio emission of PKS\,1510-089 has a core-jet structure, as pointed
out by previous observations \cite{homan02,jorstad05}. The radio
emission is dominated by the core component C from which the jet
emerges forming an angle of -28$^{\circ}$. The jet emission is not
straight, but it slightly bends at about 10 parsec from the core. Deep
multi-epoch polarimetric VLBI observations at 8.4 GHz (Fig. 1b)
point out the
presence of a low-surface brightness tail visible up to 125 pc. The
emission from the jet is mainly from the first 25 pc, where various
knots ejected from the core at different times can be identified and
followed throughout several observing epochs. For example, by
comparing 15-GHz VLBA observations from the multi-epoch MOJAVE
programme it is possible to identify different jet knots and monitor
their evolution (see Section 3). 
At 8.4 GHz both the core and the jet are polarized in
all the observing epochs.  

\begin{figure}
\begin{center}
\includegraphics{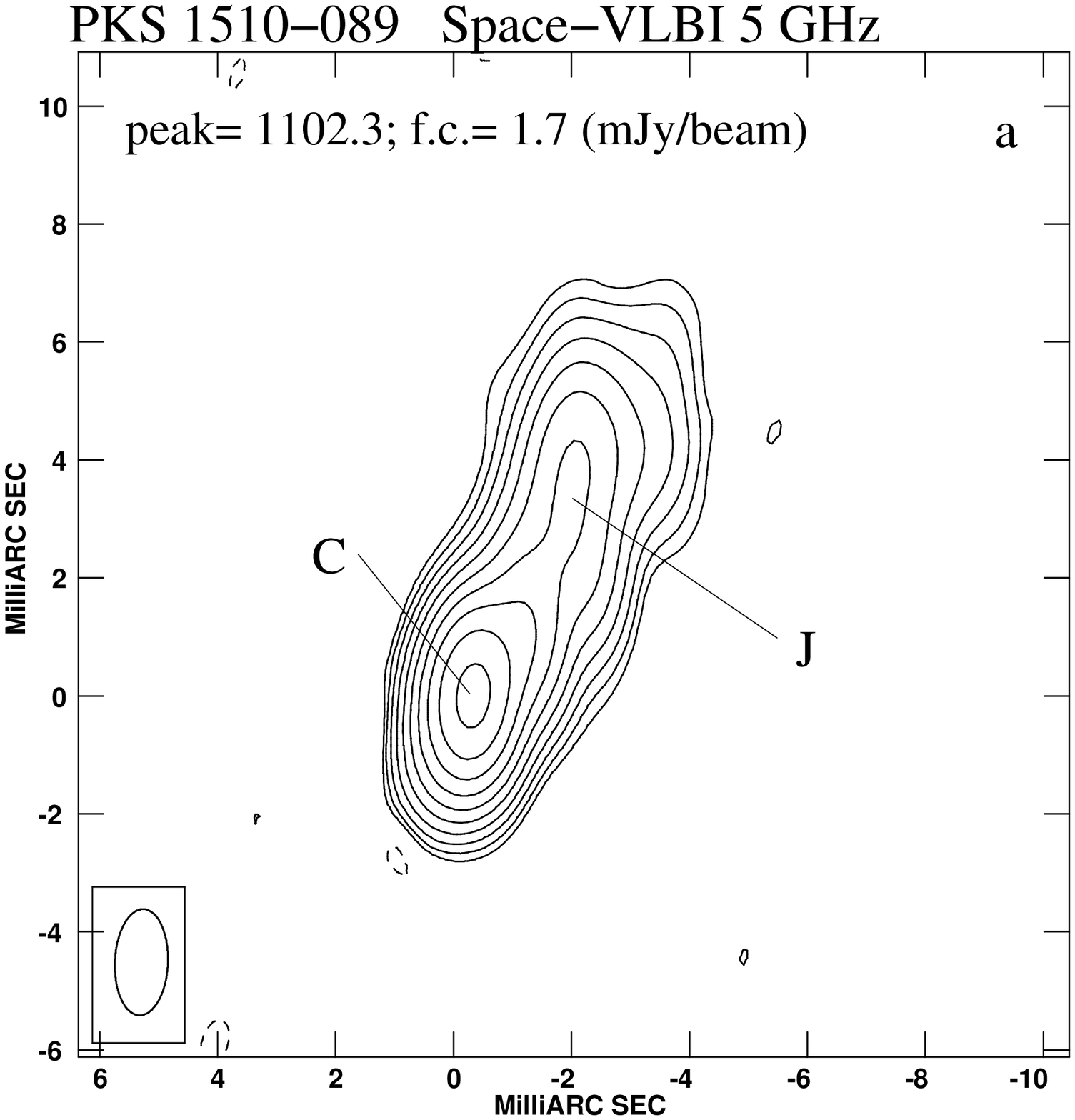}
\includegraphics{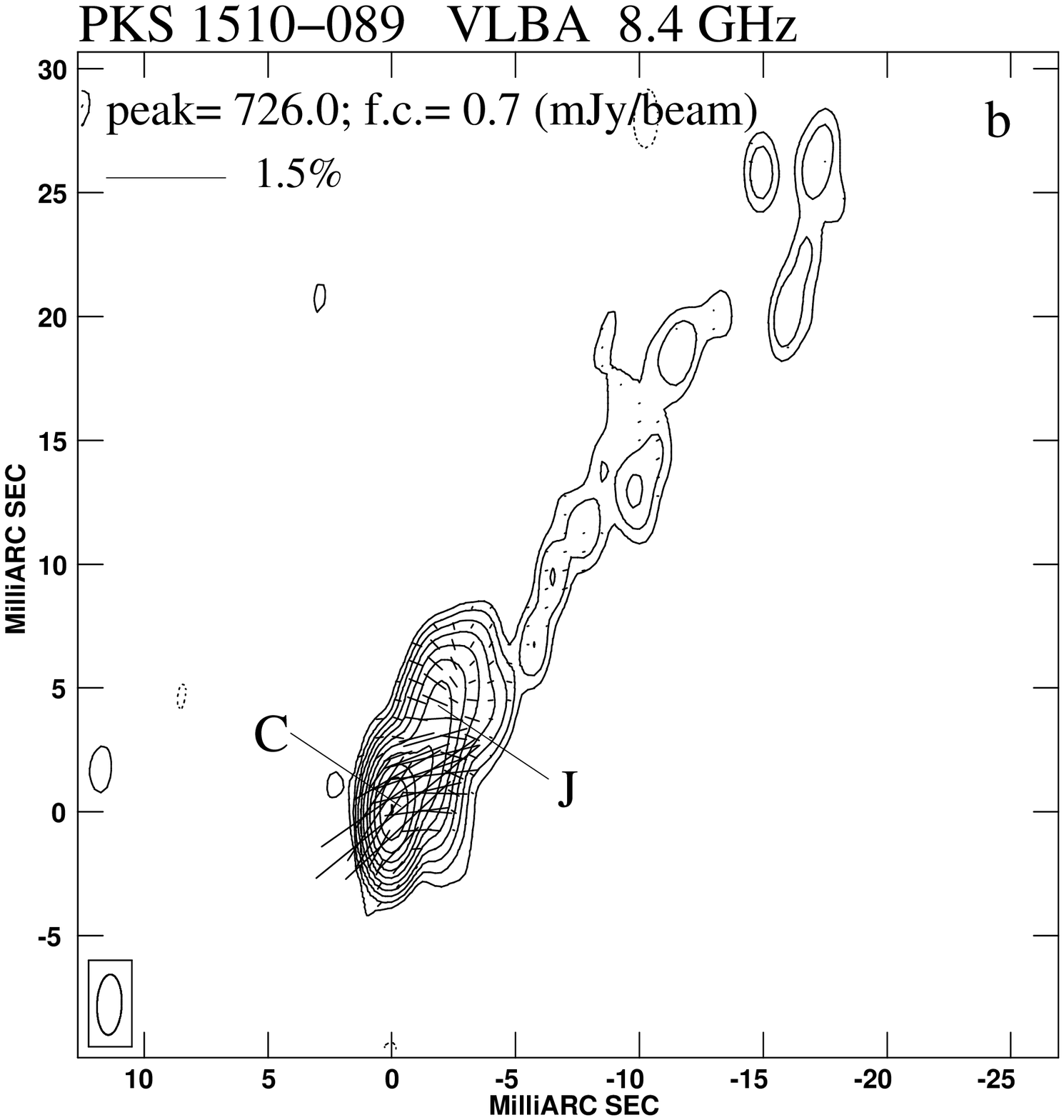}
\vspace{5cm}
\caption{5-GHz Space-VLBI ({\it left}) and 8.4-GHz VLBA ({\it right})
  images of PKS\,1510-089. The first contour (f.c.) level corresponds
  to three time the off-source noise level. Contours increase by a
  factor 2. Adapted from \cite{orienti10}.}
\end{center}
\end{figure}

\section{Proper motion}

Multi-epoch analysis of the pc-scale morphology shows an evolution of
the source structure: non-stationary jet components are ejected at
different times and their changes can be followed by comparing
observations carried out after short time intervals. From the
comparison of the three-epoch data at 8.4 GHz obtained between 1999
and 2001, we found that component
J is moving away from the central component C with a highly
superluminal apparent separation velocity $\beta$=15.9$\pm$0.3
\cite{orienti10}. To investigate the possible connection between
ejection of new jet components and $\gamma$-ray flares, we analysed
multi-epoch 15-GHz VLBA data spanning a time interval from 2007 and
2010, i.e. in the AGILE and {\it Fermi} era. During this time interval
we could follow the evolution of three knots with apparent velocity
between 16$c$ and 19.5$c$. In two cases 
regression extrapolation suggests that the
ejection of the knot occurred very close in time with a strong
$\gamma$-ray flare, as also found by \cite{marscher10}, 
suggesting a connection between the radio and
$\gamma$-ray emission.

\section{Flux density variability and polarization}

The analysis of the light-curve of PKS\,1510-089 shows strong flux
density variability where periods in which the source is in a low
state are alternated with flares \cite{venturi01}. The pc-scale
resolution of the MOJAVE data allows us to separate the contribution
of the core from that of the jet structure. The comparison between the
radio variability and polarimetric information 
of the core and the episodes of enhanced
$\gamma$-ray luminosity does not provide a clear picture. Indeed, the
high $\gamma$-ray activity states occurred in September 2007
\cite{pucella08} and September 2008 \cite{tramacere08} seem to follow
a radio flare, while the enhanced activity observed in March 2009
\cite{dammando10} seems to coincide with a low state in the radio
band (Fig. 2a). 
If we consider the polarization properties, it seems that the
polarization percentage does not correlate with the $\gamma$-ray and
total radio flux density variability (Fig. 2b). On the other hand, the
polarization angle (Fig. 2c) shows an abrupt change of about 75$^{\circ}$ close
in time with the strong increment in both radio and $\gamma$-ray
emission \cite{orienti10}. The fact that such a rotation of the polarization
angle occurs close in time with the ejection of a superluminal
component may indicate that the new component is highly polarized
with an orientation of the electric vectors that is very different
from that of the core.\\

\begin{figure}
\begin{center}
\includegraphics{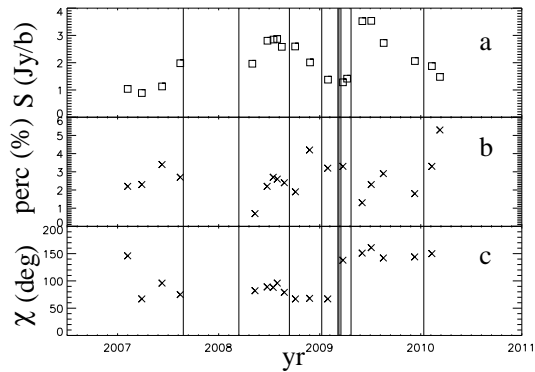}
\vspace{4cm}
\caption{Multi-epoch 15-GHz flux density ({\it a}), polarization
  percentage ({\it b}), and polarization angle ({\it c}) of the core
  component of PKS\,1510-089. Adapted from \cite{orienti10}.}
\end{center}
\end{figure}

\section{Summary}
We have presented multi-epoch high spatial resolution VLBI and
Space-VLBI observations of the core-jet FSRQ PKS\,1510-089. The
comparison of the pc-scale radio structure observed at different
epochs shows the presence of non-stationary jet components moving at
highly superluminal velocity. For two of these knots, the time of zero
separation from the core occurs close to an episode of enhanced
$\gamma$-ray activity. After these episodes, changes in the radio flux
density and in polarization properties have been observed, and they
may be due to the ejected component, spatially unresolved from the
core region, possessing a different orientation of the magnetic field
with respect to the core.  \\

\end{document}